\documentclass[nameyear]{elsarticle}
\usepackage{lineno,hyperref}
\modulolinenumbers[5]
\journal{Journal of \LaTeX\ Templates}

\usepackage[varg]{txfonts}
\usepackage{times}

\usepackage{multirow,graphicx,bm,amssymb}
\usepackage[normalem]{ulem}
\usepackage{soul}
\usepackage{epsfig}
\usepackage{natbib}
\usepackage{psfrag}

\def\apj{ApJ}
\def\apjl{ApJ}
\def\apss{Ap\&SS}
\def\aap{A\&A}
\def\mnras{MNRAS}
\def\na{New A}
\def\ssr{Space~Sci.~Rev.}
\def\nat{Nature}
\def\physrep{Phys.~Rep.}

\newcommand{\msun}{\mbox{$M_\odot$}}

\def\be{\begin{eqnarray}}
\def\ee{\end{eqnarray}}
\def\bi{\begin{itemize}}
\def\ei{\end{itemize}}
\def\lsim{\mathrel{\rlap{\lower3pt\hbox{\hskip1pt$\sim$}}
     \raise1pt\hbox{$<$}}} 
\def\gsim{\mathrel{\rlap{\lower3pt\hbox{\hskip1pt$\sim$}}
     \raise1pt\hbox{$>$}}} 

\begin{document}

\begin{frontmatter}

\title{\st{Two-}Three-peak GRBs and their implications for central engines}

\author{Enrique Moreno M\'endez}
\address{Instituto de Astronom\'ia, Universidad Nacional Aut\'onoma de M\'exico, Circuito Exterior, Ciudad Universitaria, Apartado Postal 70-543, 04510, D.F., M\'exico.\\ 
Universidad An\'ahuac M\'exico Sur, Av. de las Torres 131, 01780, D.F., M\'exico.}
\ead{enriquemm@astro.unam.mx}

\author{Nissim Fraija}
\address{Instituto de Astronom\'ia, Universidad Nacional Aut\'onoma de M\'exico, Circuito Exterior, Ciudad Universitaria, Apartado Postal 70-543, 04510, D.F., M\'exico.}
\ead{nifraija@astro.unam.mx}

\author{Barbara Patricelli}
\address{Instituto de Astronom\'ia, Universidad Nacional Aut\'onoma de M\'exico, Circuito Exterior, Ciudad Universitaria, Apartado Postal 70-543, 04510, D.F., M\'exico.\\
Universit\`a di Pisa, I-56127 Pisa, Italy and INFN, Sezione di Pisa, I-56127 Pisa, Italy
}
\ead{barbara.patricelli@pi.infn.it}

\begin{abstract}
GRB 110709B presented a peculiar three-peak lightcurve; this burst twice triggered the BAT detector onboard {\it Swift}.  
The two triggers were separated by $\sim 10$ minutes.  
In order to explain such an event, we unify into a single description the millisecond (ms) protomagnetar and the collapsar central-engine models.
We find that such a scenario could produce GRBs with three peaks.  
One for the ms-protomagnetar stage, a second one for the BH-formation event and a third one for the collapsar phase.
We show that the three peaks for GRB 110709B originate from different phases of the same collapsing object.
We estimate the energies and timescales of the different episodes of this burst using our model and compare with previous results  as well as with a reanalysis we perform on the data.  We show that not only the light curve, but also the photon index evolution and the delay between the prompt emission and the afterglow of the second central-engine activity phase point towards a model like the one proposed here.
We find that, with reasonable assumptions, our model correctly describes the activity in GRB 110709B.  We further suggest careful study of future GRBs lightcurves which may help show the validity of our model.
If our model is correct, this would be the first time that the formation of a BH from a core-collapse event is observed unimpededly.
\end{abstract}


\end{frontmatter}


\section{Introduction}
\label{sec-Intro}

The Collapsar model for long gamma-ray bursts \cite[GRBs;][]{1993ApJ...405..273W,1999ApJ...524..262M}, requires the collapse of a stellar core (up to the carbon layer) with a specific angular momentum ($a = J/M$, $J$ is the angular momentum of the collapsed core, and $M$ is its mass) of $a \gtrsim 10^{16.5}$ cm$^2$ s$^{-1}$ \citep[i.e., larger than the $a$ of the last stable circular orbit; see][]{2007Ap&SS.311..177V,2012ApJ...752...32W,2014ApJ...781....3M}.  
Under these conditions the stellar core collapses to a black hole (BH), surrounded by an accretion disk. 
This implies pre-collapse spin periods of $P_{spin}\lesssim 0.5$ day.
This severely restricts the stellar evolution prior to collapse.

One of the most likely avenues to produce a massive stellar core with such a large amount of angular momentum is by utilizing the angular momentum stored in the orbital period of a binary \citep{1998ApJ...494L..45P,2000NewA....5..191B,2002ApJ...575..996L,2003ARep...47..386T,2004MNRAS.348.1215I,2004ApJ...607L..17P,2007ARep...51..308B,2007Ap&SS.311..177V}.  
Making use of binary evolution with Case C mass transfer followed by a common envelope phase \citep[as done in][]{2002ApJ...575..996L,2007ApJ...671L..41B,2008ApJ...685.1063B,2011ApJ...727...29M} binaries with a massive He star and a $\sim 1 \msun$ companion may evolve in to an orbital period of less than a day.
This allows to spin up the He star through tidal interaction and provides the required $a \gtrsim 10^{16.5}$ cm$^2$ s$^{-1}$ to its C core.
This evolutionary method correctly predicted the spins of 4 BHs in binaries \citep[see][and refs. therein]{2014ApJ...781....3M}.
It also underpredicts the spins of BHs in high-mass X-ray binaries (HMXBs); however those may be explained through hypercritically accreting material (e.g., coming from a wind Roche lobe overflow) after the formation of the BH \citep[see][and Moreno M\'endez \& Cantiello, submitted]{2008ApJ...689L...9M,2011MNRAS.413..183M}

Alternatively, \citet{1992ApJ...392L...9D} and later works \citep[e.g.][]{2011MNRAS.413.2031M} have proposed utilizing the spin down of ms protomagnetars as the central engines for long GRBs.  
These engines have much less immediately available rotational and binding energy when compared to collapsars ($\sim10^{54}$ erg vs $\sim10^{52}$ erg)
as the mass and spin of the compact object are both smaller.
Hence, it is likely that they may trigger potentially less energetic GRBs/HNe (hypernovae).
On the other hand, 
they may be much more common as the progenitor stars could be considerably less massive and, hence, much more abundant.
It is likely that, for these engines to work, they must be much more efficient in their energy conversion.
Nonetheless, it is still necessary for the progenitor of ms magnetars to rotate extremely rapidly prior to the collapse, both, to explain its ms rotation and, perhaps even, to amplify the magnetic field to magnetar range.

Most commonly, GRBs have a single episode in the prompt phase with some (s to ms) structure, nonetheless, several GRBs show quiescent times \citep[e.g.,][]{2001MNRAS.320L..25R}.
Some of these may show a two-peak structure.
Many scenarios in terms of GRB jet composition and emission processes have been widely discussed to explain the structure of lightcurves with two salient peaks.   
Among them there are GRB 980923,  GRB 990123,  GRB 041219A and GRB 110731A 
GRB 030329 
\citep{2012ApJ...755..140G,2011arXiv1110.6421F, 2012ApJ...751...33F,2015ApJ...804..105F}.  
GRBs 990123, 041219A and 980923 have been described as to have early emission produced by the forward shock at its early stage when it propagates into the pre-accelerated and pair-loaded environment  \citep[][and refs. therein]{2012ApJ...751...33F}. 

It is worth noting that \citet{2008MNRAS.383.1397L} have proposed a model capable of explaining a two-peak GRB before.  Nonetheless, their model depends on a Spinar (a very rapidly rotating star) which collapses, the dissipation produces a first burst, which is later followed by a second burst from a central engine.  We do not consider such a spinar stage in our model. 

In this paper we study the possibility, previously suggested in \citet{2012ApJ...748..132Z}, that GRB 110709B was the result of combination of factors that allowed both mechanisms, ms magnetar and collapsar, to play a part in this transient event.
It is beyond the scope of this work, however, to attempt any numerical simulations of central engines and/or core-collapse SNe; we refer the reader to recent literature on simulations of collapsars using Blandford-Znajek \citep{2008MNRAS.385L..28B, 2009MNRAS.397.1153K, 2010MNRAS.401.1644B} or not \citep{2008ApJ...673L..43D}, as well as those using ms magnetars.  
In section~\ref{sec-Model} we briefly describe the ms-protomagnetar/Collapsar scenario and suggest a list of observables were this model correct.
In section~\ref{sec-Results} we show a short reanalysis of the Swift BAT data and show that it provides further evidence suggesting the existence of a second-stage central engine, probably the result of a collapse into a BH and the switch-on of a Blandford-Znajek \citep[or BZ;][]{1977MNRAS.179..433B} engine.
In section~\ref{sec-Discussion} we discuss the new insights from the reanalysis and their implications for the engine of GRB 110709B.
Finally, in section~\ref{sec-Concl} we discuss the implications for the models of central engines, the jets and their interaction with the surrounding stellar material which may (or, as after the first episode of GRB 110907B, may not) produce the accompanying HN.

\section{Model}
\label{sec-Model}

GRB 110709B presents a light curve that shows three distinct episodes.
\citet{2012ApJ...748..132Z} and \citet{2013A&A...551A.133P} treat the second and third episodes as a single event.  
However, as we shall discuss in section~\ref{sec-Results}, by reanalyzing the data we show that they are two different events.
Thus, hereafter we will refer to these as episodes 1, 2 and 3.

To explain the three episodes of GRB 110709B as the collapse of a single star (although, in a binary) we require a model that can differentiate these episodes and explain their time separation (of $\sim 10$ and $1$ minutes, respectively).
Thus, we propose a star which develops a centrifugally-supported, massive, $\sim 3.5 \msun$ core which collapses in two stages.  
The first stage (1.5 to 2.5 $\msun$) falls in a dynamical timescale and produces a ms magnetar which provides the central engine for the first episode of the GRB.  
The remaining $1$ to  $2 \msun$ have too much angular momentum and fall in through an accretion disk in a viscous timescale.
This results in a quiescent period, once the magnetar slows down (after $\sim 100$ seconds), through which the magnetar accretes another 1 to $2 \msun$ from the remaining core.
The second burst stage involves the collapse of the compact star into a Kerr BH releasing in a few seconds the binding energy which produces a second event.
Lastly, the BH switches on as a BZ central engine and produces the third episode.
We now proceed to detail these events.

\subsection{Act I:  Magnetar stage}
\label{sec-Magnetar}

Since we expect to rapidly ($\tau \lesssim 10$ minutes) form a Kerr BH, we consider a collapsing $3.5 \msun$ core ($R_{Fe}\simeq 10^9$ cm). The free-fall timescale of such an object is of the order of 
\be
\tau_{ff} = \frac{\pi}{2} \sqrt{\frac{R_{Fe}^3}{GM_{Fe}}} \simeq \frac{3 \pi}{4} \;
\left(\frac{R}{10^9 {\rm cm}}\right)^{3/2}\left(\frac{M}{3.5\msun}\right)^{-1/2} {\rm s}. \label{eq:Kepler}
\ee
Instead, if the material is centrifugally supported and falls through an accretion disk, with viscosity $\alpha \sim 0.01$ \citep{2005ApJ...632..421L} and thickness $r/h \sim 1$ (since it is still forming from a collapsing stellar hot core we expect it to be quite thick), the viscous timescale, $\tau_v$, is
\be
\tau_v \simeq \left(\frac{r}{h}\right)^2\frac{4\tau_{ff}}{\alpha} \simeq 900 {\rm s},
\ee
which is comparable to the delay of the second episode in GRB 110709B.

The (Newtonian\footnote{For relativistic corrections to the moment of inertia see, e.g., \citet{2010arXiv1012.3208L}}) rotational kinetic energy of a protoneutron star (PNS; or a neutron star, NS) is then of the order of
\be
E_k = \frac{1}{2}I \Omega^2 = \left(\frac{1}{2}\right)\left(\frac{2}{5}\right) \frac{G M^2}{R} 
\simeq  133 \left(\frac{k_s}{2/5}\right) \left(\frac{M}{\msun}\right)^2 \left(\frac{10^6 {\rm cm}}{R}\right) {\rm B},
\ee
where, a Bethe, $1$ B $= 10^{51}$ erg; $k_s$ is the coefficient of the moment of inertia (which is 2/5 for a perfect sphere).  $\Omega$ is the resulting angular velocity of the PNS or NS after collapse and is of the order $\Omega \lesssim 10^{4}$ s$^{-1}$ which is close to its break-up speed.
Thus the slowdown of the compact object could power a GRB.
Choosing a $2 \msun$ protomagnetar of $R_{pns} \sim 50$ km (and spinning at break-up speed) will have 42.5 B of available energy; if one waits some tens of seconds for the neutrinos to cool down the PNS, the radius of the NS will be $R_{ns} \sim 10$ km, thus part of the binding energy would go into spinning up the NS and the available (rotational kinetic) energy will be $\sim 213$ B instead.

The rotational energy of the (proto)magnetar can be tapped through a torque exerted by the magnetic dipole of the magnetic field \citep[$B$;][]{1992Natur.357..472U}.  The power, or luminosity, for this process can be estimated from
\be
\dot{E}_k \simeq \frac{2}{3}\frac{B^2 R^6 \Omega^4}{c^3} \simeq 2.2 \left(\frac{B}{3 \times 10^{14} {\rm G}}\right)^2\left(\frac{\Omega}{10^{3} {\rm s}^{-1}}\right)^4 {\rm B\; s}^{-1},
\ee
where $R$ is the radius of the magnetic dipole and $\Omega$ is the angular velocity.  

After a few tens of seconds, the ms-magnetar engine slows down and the jets are shut down.
At this point, matter accumulated in the accretion disk, and likely held there by propeller effect, will start streaming down onto the magnetar, increasing its mass and burying its magnetic field.

It is important that the supernova (SN) shock does not succeed at dismantling the core of the star, or else, there will be no sequel GRB.
At most, the shock (and its reenergizing by the ms magnetar) could succeed at bouncing out the core to a few $10^9$ cm to $10^{10}$ cm.  Then, the material would fall back and form a new accretion disk.  This would provide the ingredients for the second  central engine a few minutes later.

During this new accretion stage, the material forming the magnetar may be highly magnetized in the interior, but its exterior field may be low as the field may remain buried for an Ohmic timescale (which is much longer than the dynamical timescale).
As the material from the accretion disk piles up on the surface of the magnetar it transfers angular momentum back onto its surface (and inwards, as the Alfv\'en timescale is extremely short).
Even if the bulk of the magnetic field of the magnetar is buried by this new material, magneto-rotational instabilities (MRIs) and dynamos may rebuild a substantial magnetic field in the accretion disk. 
The high temperature, high internal magnetic field and high rotation ration rate of the magnetar may keep it for a few tens of seconds from collapsing, however its high mass, well above the typical threshold for a NS, will eventually overcome the strong pressure and a Kerr BH will be formed.

\subsection{Act II:  BH formation}
\label{sec-BH}

If we assume the binding energy released during the conversion from NS to BH is around 1 to $10 \%$ of the total mass, (we have assumed that even for a stiff equation of state, a large magnetic field, and almost Keplerian rotation this should occur below $M_{NS} \sim3.5 \msun$) then the total energy released in this event should be of the order of 
\be
E_T = k_G M_{NS}c^2 =  630 \left(\frac{k_G}{0.1}\right)\left(\frac{M_{NS}}{3.5\msun}\right) {\rm B},
\ee
where $k_G$ is the fraction of mass converted to energy.
Under {\it normal} SN conditions, over $99 \%$ of the energy released leaves the star as neutrinos, without further interaction.
For SN 1987A \citep{2014MNRAS.442..239F} the kinetic energy was around $E_{kin} \sim 1$ B and the total energy released (in neutrinos) was $E_T \sim 300$ B.  Assuming somewhat larger efficiency (larger density and temperature, thus, larger neutrino cross section), and considering we have more mass in the compact object we estimate the kinetic energy to be a few times larger than in SN 1987A, i.e., $E_{kin}  \sim 5$ B, which coincides with the observations for GRB 110709B.

The first GRB episode drilled holes through the star.
However, maybe due to a large lateral density gradient (along the polar angle), the cocoons
\footnote{As described in \cite{2002MNRAS.337.1349R}, as the GRB jet propagates inside the star, before it breaks out, energy is deposited on the surrounding material.  This material propagates perpendicular to the jet, thus creating a cavity or cocoon which expands laterally and may eventually blow the star apart in a hypernova explosion.} 
do not disrupt the star.
Still, some fallback or low-specific-angular-momentum material will likely start accumulating along these paths.
We expect the SN shockwave to be directed up these partially clogged nozzles as the resistance is much lower in these directions.
Given the nature of this collapse, where nuclear matter collapses into a BH (similar to the case of a short GRB, where a NS-NS or a BH-NS system merges) it is likely that this event will produce a substantial flux of neutrinos (and gravitational waves), hence, sparing the rest of the star from a dangerously energetic shockwave which could rapidly dismantle it.
Otherwise, the central engine will not be in place to produce the third episode of GRB 110709B, which lasts in excess of 250 seconds.

Given the channeling of all the kinetic energy and the fact that the material that forms the collapsing magnetar is denser than that in the accretion disk,   
a quasi-thermalized, sharp, hard-X-ray signal would be expected. 
Thus, we do not expect this explosion to be SN-like, but instead more like a harder-GRB signal.
Furthermore, the channeled shockwave and its echoes may further develop into a train of shockwaves (much like in a GRB) that will, likely, interact between themselves far away from the star.
Similar to the scenario described by \citet{2001MNRAS.320L..25R} and \citet{2001MNRAS.324.1147R}. 
\subsection{Act III:  BZ engine}
\label{sec-BZ}

The second episode may again push the accretion disk outwards, allowing it to fall back in a timescale of a few tens of seconds.
As further material and angular momentum are accreted into the BH, and as the magnetic field intensifies, a BZ  central engine will replace the ms-magnetar engine.  It is likely that the neutrino annihilation may also play a role in powering this third episode of GRB 110709B as well as in keeping a steep lateral density gradient during the prolonged quiescent phase.
In principle, this last event is (almost) unimpeded, thus, all the energy usually required to drill through the star may simply go into the jet. 
Hence, this new GRB will be observable for as long as the central engine remains in place.  Something, likely, not seen in any previous GRB.

Following the analysis performed by \citet{2000PhR...325...83L} we estimate the relevant quantities for a BZ central engine.  First, we obtain the available energy to the BZ process:
\be
E_{BZ} = 1,800 \; \epsilon_\Omega f(a_\star) \left(\frac{M}{\msun}\right) {\rm B},
\ee
where $a_\star \equiv Jc/(GM^2) = a/M$, 
\be
f(a_\star) = 1 - \sqrt{\frac{1}{2}\left(1+\sqrt{1-a_\star^2}\right)},
\ee
and $\epsilon_\Omega = 0.5$, the maximum efficiency for energy extraction. 
Next, we calculate the power or luminosity of the BZ central engine (in Bethes per second) which depends not only on the spin of the BH ($a_\star$) but also on the magnetic field permeating the region \citep[see appendix D in][]{2000PhR...325...83L} :
\be
P_{BZ} \simeq 0.17 \;a_\star^2 \left(\frac{B}{10^{15}{\rm G}}\right)^2\left(\frac{M}{\msun}\right)^2 {\rm B \; s}^{-1}.
\ee

Early during the BZ-engine regime, the mass of the BH should be above $3.5\msun$ (which it had when it was formed).  
This is also necessary as the magnetar switched off because of its low spin.
Thus, the accreted material must resupply the central compact object with angular momentum.  
Similarly, the magnetic field will likely be regenerated during this accretion phase \citep[the energetic cost is extremely low; see, e.g.,][]{2014ApJ...781....3M}. 
With these considerations in mind, we estimate then $M_{BH} \sim 4 \msun$, $a_\star \sim 0.7$ and $B \sim 10^{15}$ G.
Utilizing these numbers then the total energy available to the central engine is $E_{BZ} \sim 260$ B.
The luminosity is then $P_{BZ} \sim 1.3$ B s$^{-1}$.  
Hence, $T_{BZ} \sim E_{BZ} / P_{BZ} \sim 200$ s.
These numbers reflect fairly well those observed for GRB 110709B, especially if we consider that part of the energy may be lost to neutrinos and GWs.


\subsection{Regarding Afterglows}
\label{sec-AGs}

\begin{figure}
\begin{center}
\includegraphics[scale=1,angle=0]{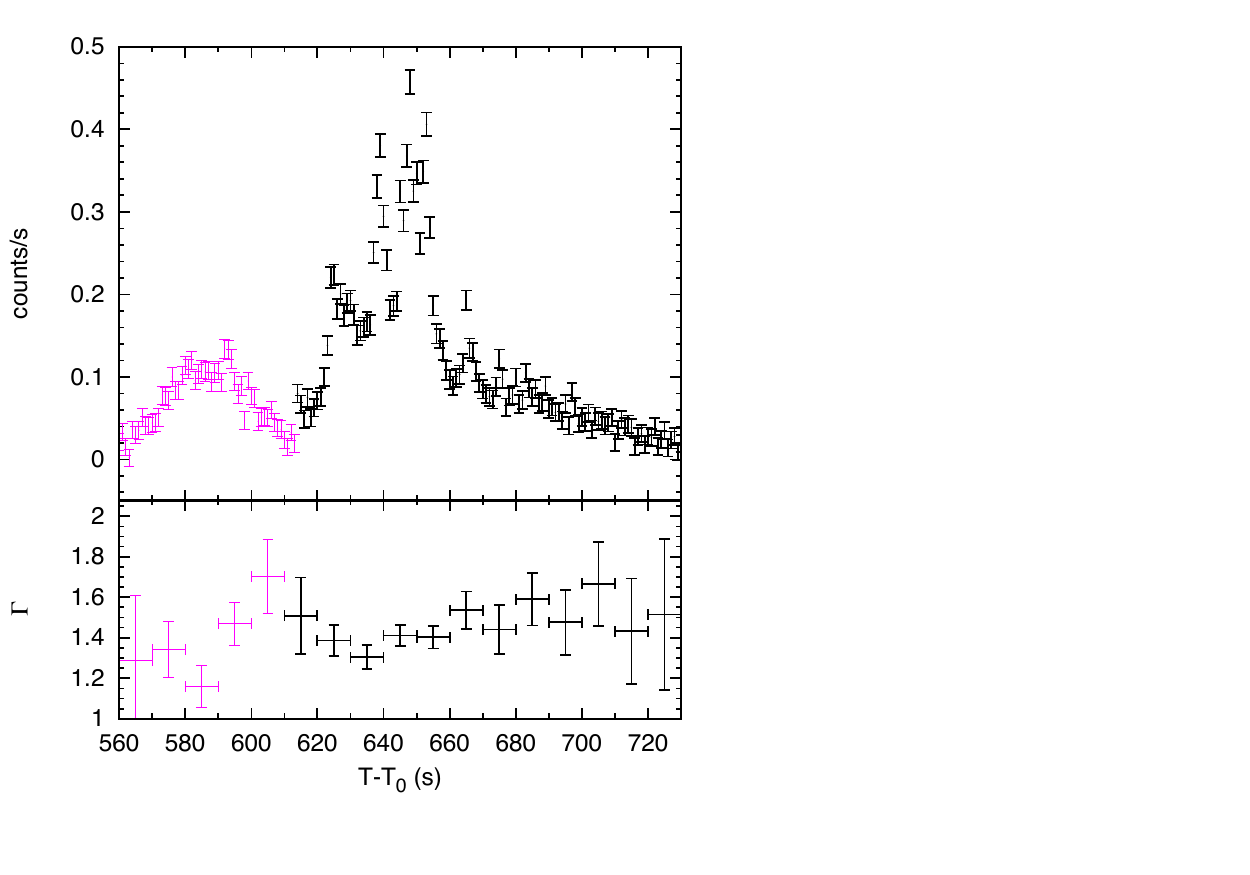}
\caption{BAT count rates (upper panel) and photon index evolution (lower panel) during episodes 2 (in magenta) and 3 (in black) of GRB 110709B. The spectral model is a simple power law. The time is relative to the trigger time of episode 1, of 331940612 s (in MET seconds).}
\label{fig:LC}
\end{center}
\end{figure}

The fireball model \citep[see][for recent reviews]{2006RPPh...69.2259M, 2004IJMPA..19.2385Z} satisfactorily explains GRBs and their afterglows (AGs).  This model predicts  an  expanding  ultrarelativistic shell that moves into the external surrounding medium.  
As this ejecta sweeps and accumulates circumstellar material on its head, it slows down and the relativistic beaming stops.
For instance, the deceleration radius and timescale can be written as
\begin{equation}
R_{d}=\biggl(\frac{3}{4\pi\,m_p}\biggr)^{1/3}\,\Gamma^{-2/3}\,\eta^{-1/3}\,E^{1/3}\, ,
\end{equation}
and
\begin{equation}\label{eq:Tdec}
  t_{d}=\biggl(\frac{3}{32\pi\,m_p}\biggr)^{1/3}\,(1+z)\,\Gamma^{-8/3}\,\eta^{-1/3}\,E^{1/3}\,,
\end{equation}
respectively, where $E$ is the energy, $\Gamma$ is the bulk Lorentz factor, $\eta$ is the interstellar medium (ISM) density and $m_p$ is the proton mass.

If the three episodes observed in GRB 110709B have a common progenitor, it would be expected that the AG produced by the last episode takes longer to occur than the previous ones.
This is a consequence of material being swept out of its path onto larger radii by previous episodes.
Furthermore, if the later episodes posess AGs, it is necessary that the earlier ones also had them.

According to the standard relativistic fireball model, the forward
shock accelerates electrons up to relativistic energies.
This generates magnetic fields through the first-order Fermi mechanism or through electric fields associated with the Weibel instability. 
The afterglow emission is more likely to be synchrotron, thus the spectrum has a break in the fast-cooling regime resulting in a spectral index of $\alpha = 1.5$ \citep{1998ApJ...497L..17S,1999ApJ...524L..47G}.




\section{Observations and Data Analysis}
\label{sec-Results}

\begin{figure}
\resizebox{8.8cm}{!}{\includegraphics[angle=0]{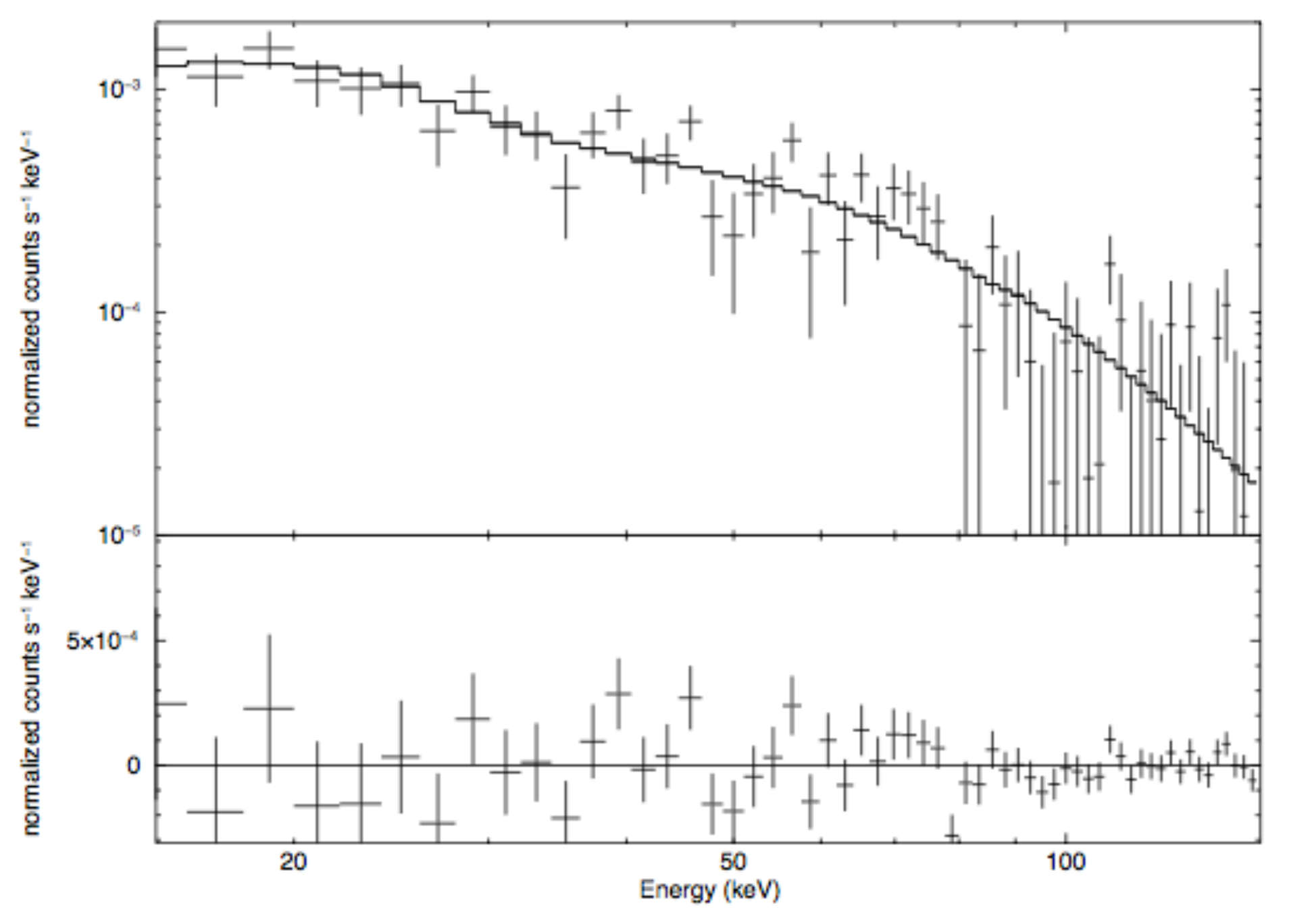}}
\caption{Fit of the Swift/BAT spectrum between 835 s and 865 s after the trigger time of episode 1 with a PL model. The lower panel shows the residuals.}
\label{fig:AG}
\end{figure}

GRB 110709B triggered the Burst Alert Telescope (BAT, \citealp{2005SSRv..120..143B}) onboard the Swift satellite at 21:32:39 UT on 2011 July 9 \citep{2011GCN..12144...1C}. 
This episode extended up to 55 s after the trigger \citep{2012ApJ...748..132Z}. 
Interestingly, there was a second BAT trigger at 21:43:25 UT on 2011 July 9, $\sim$ 11 minutes after the first trigger time. 
The emission period extended up to 865 s after the first trigger \citep{2011GCN..12144...1C}; this period shows a first bump (episode 2), beginning at $\sim$ 550 s after the first trigger and lasting about 60 s (ending at 610 s), followed by a second multi-peaked bump (episode 3, from 610 s to 750 s; see Fig. 1) of longer duration.  

\citet{2012ApJ...748..132Z} performed a time-resolved spectral analysis of this second emission (episodes 2 and 3), dividing its time period in time slices of 50 s or more; they show that the spectra can be fitted with cutoff power laws and that there is a strong hard-to-soft spectral evolution. 
However, their choice of the time intervals is not suitable to look at the evolution of the spectral parameters during episode 2, as the corresponding time slice covers its whole duration. 
Therefore, we performed a more detailed analysis of the spectra of episodes 2 and 3, by considering sub-intervals of 10 s each. 
We processed the \emph{Swift}/BAT data using the standard FTOOLS package (Heasoft, version 6.15). 
We analysed the spectra using two different spectral models: BB+PL and PL. We found that all the spectra are well modeled with PL, while the BB+PL can be discarded, as it is not well constrained. 
The lower panel in Fig. 1 shows the time evolution of the photon index of the PL model: it can be seen that a discontinuity in the hard-to-soft evolution comes out at the beginning of episode 3. 
This could suggest a different emission mechanism for episodes 2 and 3.

We obtain that the BAT band (15-150) keV fluences of the first (from $-28$ to $55$ s from first BAT trigger), second ($550$ to $610$ s from first BAT trigger) and third ($610$ to $750$ s from first BAT trigger) episodes are $9.54^{+0.11}_{-0.16} \times 10^{-6}$ erg cm$^{-2}$, $2.31^{+0.06}_{-0.05} \times 10^{-6}$ erg cm$^{-2}$ and $8.81^{+0.09}_{-0.12} \times 10^{-6}$ erg cm$^{-2}$, respectively.

We also analyze the spectrum between 835 s and 865 s corresponding to the last peak observed in the BAT light curve \citep[see][]{2012ApJ...748..132Z}, tenths of seconds after the end of episode 3. We interpret this peak as the AG of episode 3 (see sec. 4).
We find that the BAT spectrum (see Fig.~\ref{fig:AG}) is well modeled with a PL with photon index $\alpha_3 = 1.59^{+0.15}_{-0.14}$ (where the subindex denotes that this belongs to the AG of episode 3).
We summarize the observed and estimated (from our model) quantities in table~\ref{Tab:ExpVsTeo}.




\section{Discussion}
\label{sec-Discussion}

\begin{table*}       
\begin{center}
\caption[]{Here we show the observed versus the estimated quantities for GRB 110709B. Derived quantities are obtained from other values also listed in the table.  All values come from either \citet{2012ApJ...748..132Z} or are estimated with our model; our observed energies use the redshift from \citet{2013A&A...551A.133P}.}\label{Tab:ExpVsTeo}
\begin{minipage}{126mm}
\begin{tabular}{lllccccccc}
\hline \hline
\multicolumn{10}{c}{GRB 110709B}\\
\hline \hline
 & &  &  E  &   F  &  T$_{AG}$ &  P  &  $\alpha$ & $\Gamma$ & $\eta$ \\ 
&&&[B]& [$10^{-6}$ erg cm$^{-2}$] & [s] & [B s$^{-1}$] & & & [cm$^{-3}$]\vspace{1mm}\\      
\hline
&& E1 & 20.0 & $9.54^{+0.11}_{-0.16}$ & $\sim$ 40 & -- & $1.5^{+0.05}_{-0.05}$ & -- & -- \vspace{1mm}\\ 
&& E2 & 4.85 & $2.31^{+0.06}_{-0.05}$ & -- & -- &  --  & -- & -- \vspace{1mm}\\
\rotatebox{90}{\parbox{2mm}{\multirow{3}{*}{\hspace{-1mm}\footnotesize{Observed}}}}&& E3 & 18.5 & $8.81^{+0.09}_{-0.12}$ & $\sim$ 245 & -- & $1.59^{+0.15}_{-0.14}$ & -- &  -- \vspace{1mm}\\
\hline
&& E1 &  133 &  --  &     40    & 2.2 & 1.5 & 160 &     3    \vspace{1mm}\\ 
&& E2 &  630 &  --  &     --    &  -- &  -- &  -- &     --   \vspace{1mm}\\
\rotatebox{90}{\parbox{2mm}{\multirow{3}{*}{\hspace{-1.5mm}\footnotesize{Estimated}}}}&& E3 &  260 &  --  &    280    & 1.3 & 1.5 & 225 & $10^{-3}$ \vspace{1mm}\\
\hline\hline
\end{tabular}
\end{minipage}
\end{center}
\end{table*}

The model described in sec.~\ref{sec-Model} predicts the existence of a second peak in the light curve.  This second peak can easily be observed in most energy bands shown by \citet{2012ApJ...748..132Z}.
Nonetheless, as mentioned in sec.~\ref{sec-Results}, their analysis  assumes that episodes 2 and 3 are a single one, similar to other GRBs.  
For GRB 110709B, a more detailed analysis of the spectrum allows to differentiate three different episodes.
Furthermore, our model predicts episode 2 to be harder in spectra than episodes 1 and 3.
Unfortunately, data from higher energy band detectors does not seem to be available \citep{2011GCN..12135...1G}, but the data does suggest this may be possible.

Our model is also consistent with the, apparently controversial, observation that episode 2 cannot be explained with a thermal component in addition to the PL.
In principle episode 2 could be expected to be more SN-like, given that it is, after all, the collapse of the PNS into a BH.  
However, a SN event would disrupt the star an no third episode would be, hence, expected.
Instead, the shock produced by the collapse has to be channeled through the jet-drilled nozzles and, thus, could become a more hard GRB-like signal.

If all episodes of a multi-peak GRB are coming from the same object it should also be expected that the AG of episode 1 should appear earlier after its prompt-emission phase when compared to the AGs of the following episodes. 
By considering the values of the first episode: energy $E=20$ B, redshift $z = 0.75$ \citep[$\sim3.1$ Gpc; ][]{2013A&A...551A.133P}, ISM density  $\eta=3\, {\rm cm^{-3}}$ and Lorentz factors of $\Gamma=160$ and $\Gamma=50$, we obtain deceleration radii of $R_{d}=3.5\times 10^{16}$ cm and $R_{d}=7.5\times 10^{16}$ cm which correspond to deceleration times of $t_d \simeq 40$ s and $t_d \simeq 870$ s, respectively.

Now, taking into account the values for episode 3, energy $E = 18.5$ B, a clean ISM density  $\eta = 10^{-3}\; {\rm cm^{-3}}$ and a bulk Lorentz factor of $\Gamma = 225$, we obtain a value of deceleration time  $t_d \simeq 225$ s which is in agreement with the observations.  

The first jet propagates out of the star and into the ISM clearing a path, thus lowering $\eta$ from $3\, {\rm cm^{-3}}$ to $10^{-3}\; {\rm cm^{-3}}$.  
Hence, the jet for the next episodes will not considerably slow down until it reaches the head of the first shell, at which point it will slow down and produce its AG.
This is consistent with the observations.
\citet{2012ApJ...748..132Z} report a component modeled with a power law of spectral index $\alpha_1 = 1.55\pm0.05$ at the very end of the prompt emission, between $\sim36$ and $\sim45$ s which strongly suggests an early AG for the first episode.
Furthermore, \citet{2011GCN..12142...1D} confirm this AG with observations from XRT starting at $\sim 70$ s after the first BAT trigger. 
Instead, the AG of episode 3 may be related to the peak observed at $\sim 850$ s in the BAT and XRT lightcurves \citep[see][]{2012ApJ...748..132Z}, i.e., some $\sim 225$ s after the end of the quiescent period separating episode 1 from episode 3.  This is supported by our estimated spectral index of $\alpha_3 = 1.59^{+0.15}_{-0.14}$ (see sec.~\ref{sec-Results}). 
These $\alpha$s, along with the PL shape and the fact they are observed with BAT (XRT and/or KW) imply synchrotron radiation from AGs. 

\section{Conclusions}
\label{sec-Concl}

Although our results do not show conclusive evidence for the presence of both, the ms magnetar and Collapsar engines working one after the other in the same progenitor, they do strongly suggest that the model where both contribute to GRB 110907B is consistent and favored.
If this paradigm is confirmed in further GRBs it would reveal important information regarding the central engines. 

GRB 110709B in its first episode has an estimated fluence of $F^{(1)} = 9.54^{+0.11}_{-0.16} \times 10^{-6}$ erg cm$^{-2}$.  
In the second episode $F^{(2)} = 2.31^{+0.06}_{-0.05} \times 10^{-6}$ erg cm$^{-2}$.
And finally, in the third episode $F^{(3)} = 8.81^{+0.09}_{-0.12} \times 10^{-6}$ erg cm$^{-2}$.
This translates 
($z = 0.75$) into an isotropic energy of $E_{iso}^{(1)}\simeq 20.0$ B for the first episode, $E_{iso}^{(2)}\simeq 4.85$ B for the second one and $E_{iso}^{(3)}\simeq 18.5$ B for the third one.  Note that, if not well focused (probably to cover less than $4\pi/100$), either episode 1 or 2 would rapidly dismantle the star, thus preventing the second and/or third episodes from ever occurring.  
This is interesting as the first episode of GRB 110709B is not a low luminosity event.
In ultra-long GRBs (ulGRBs) it is clear that the star must survive the SN dissmantling by the cocoon, otherwise the central engine would be starved and the GRB would shut down at much earlier times.
However, in ulGRBs, the engines have, usually, low power as can be seen by the low luminosity of such events.
This result implies that a jet from a ms-magnetar engine, where the SN shockwave fails, may be well collimated.
It must also drill through the star rapidly enough such that the cocoon 
has little energy and cannot blow the star away in a SN explosion.
This is not completely unheard of, as e.g. shown in the list of long GRBs by  \citet{2011arXiv1104.2274H}, where a SN lightcurve seems to be nonexistent or, at best, really underluminous.

The delay of the second AG with respect to the first one, as well as with respect to the one observed in other double-peaked GRBs is interesting.
It may indicate that those other GRBs may have had a previous ms-magnetar stage which failed to produce a GRB but which opened up a channel in the structure of the progenitor star through which the BH-forming shockwave could exit and later a collapsar engine would turn on and produce a GRB.
This would produce the distinctive double-peaked lightcurve.

GRBs such as GRB 110709B, with two clear central-engine stages are capable of providing rare insights into, both, the ratio of energy that may be contained in the jet to that in the cocoon, as well as to the collapse of an overweight NS into a BH.
It is of great importance to obtain data in a wider range of energy from these transients, as this should allow to distinguish between a stage such as episode 2 in our model, where a NS turns into a BH, from activity from a Collapsar central engine, i.e., episode 3.

If the chain of events we have described here is correct, then, GRB 110709B has provided us with the first unhindered view of the formation of a BH from the core-collapse of a massive star.  
This is remarkable as this would be expected from channels such as neutrinos and/or gravitational waves, but not from photons.
Else, it can be argued that short GRBs allow this as well.  
However, in this event, the GRB-emptied out polar region of a star allows us a rare glimpse into the collapse of an overweight, accreting neutron star into a BH.

%
%
%
%
%


\section*{Acknowledgements}
EMM had support from a CONACyT fellowship.  NF is supported by a Luc Binette scholarship and the projects IG100414 and CONACyT 101958. 
We thank William Lee for useful discussions.
This research made use of NASA’s Astrophysics Data System as well as arXiv.

\bibliographystyle{elsarticle-num}



\end{document}